# Assessment of the MERS-CoV epidemic situation in the Middle East region

Chiara Poletto[1,2,3], Camille Pelat[4,5], Daniel Levy-Bruhl[6], Yazdan Yazdanpanah[4,5,7], Pierre-Yves Boelle[1,2], Vittoria Colizza[1,2,3]

[1]Inserm, U707, Paris, France
[2]UPMC Univ Paris 06, Faculté de Médecine Pierre et Marie Curie, UMR S 707, Paris, France
[3]Institute for Scientific Interchange (ISI), Turin, Italy
[4]ATIP/Avenir Inserm, U738, Paris, France
[5]Univ Paris 07, Faculté de Médecine Bichat, Paris, France
[6]Institut de Veille Sanitaire (InVS), St Maurice Cedex, 94415, France
[7] Service des Maladies Infectieuses et Tropicales, Hôpital Bichat Claude Bernard

Correspondence to : VC (vittoria.colizza@inserm.fr)

The appearance of a novel coronavirus named Middle East (ME) Respiratory Syndrome Coronavirus (MERS-CoV) has raised global public health concerns regarding the current situation and its future evolution. Here we propose an integrative maximum likelihood analysis of both cluster data in the ME region and importations in Europe to assess transmission scenario and incidence of sporadic infections. Our approach is based on a spatial-transmission model integrating mobility data worldwide and allows for variations in the zoonotic/environmental transmission and underascertainment. Maximum likelihood estimates for the ME region indicate the occurrence of a subcritical epidemic ($R = 0.50$, 95% confidence interval (CI) 0.30–0.77) associated with a 0.28 (95% CI 0.12–0.85) daily rate of sporadic introductions. Infections in the region appear to be mainly dominated by zoonotic/environmental transmissions, with possible underascertainment (95% CI of estimated to observed sporadic cases in the range 1.03–7.32). No time evolution of the situation emerges. Analyses of flight passenger data from the region indicate areas at high risk of importation. While dismissing an immediate threat for global health security, this analysis provides a baseline scenario for future reference and updates, suggests reinforced surveillance to limit underascertainment, and calls for increased alertness in high-risk areas worldwide.

**Introduction**

As of August 31, 2013, a total of 108 laboratory confirmed cases of human infection with the Middle East Respiratory Syndrome Coronavirus (MERS-CoV) have been reported to the World Health Organization (WHO) [1]. Since the emergence of the virus, a rapid coordinated response has been put in place to confront the novel epidemic emergency, through the identification and sequencing of the virus [2], the enhancement of surveillance systems, the provision of updated information on the situation and of technical guidance for the clinical management of probable infections [3-7], the identification of the possible virus reservoir [8]. There are still many uncertainties about various aspects of the outbreak, including a possible extension of the reservoir to other hosts, its full geographic extent, the path of transmission of the infection to humans and the associated risk. All these aspects call for heightened surveillance, enhanced investigations and the development and application of epidemiological methods to assess the current epidemic situation and determine its potential to spread efficiently in human population and to widely circulate at a global scale.

In such situation, statistical, mathematical and computational methods allow estimating key epidemiological parameters from available data, under various assumptions and accounting for the many uncertainties. The reproductive number, i.e. the average number of secondary cases generated by a primary case, is a key summary measure of the transmissibility of an emerging infection. A first estimation of the MERS-CoV reproductive number was based on the analysis of cluster-size data with assumed cluster partition in terms of transmission trees, highlighting the similarity of current MERS-CoV situation to the prepandemic stage of the severe acute respiratory syndrome (SARS) outbreak [9]. Next to transmissibility, an additional important aspect characterizing the epidemic remains unknown, i.e. the incidence of infection. Observed cases may indeed only represent a proportion of the current epidemic, with a majority of infections going undetected because of mild illness or asymptomatic infection. This aspect also has further relevant implications for the correct estimation of other important overall statistics (e.g. the severity of the virus) and of the risk of importation of cases in other regions of the world. Limited data may also hide important changes in the virus transmissibility related, e.g., to viral adaptations to humans that may alter its pandemic potential, thus presenting an additional challenge for the situation assessment.

Here we present an innovative integrative maximum likelihood approach to describe the current epidemic situation in the Middle East (ME) region and fill these gaps in current knowledge. We synthesize evidence from multiple sources of information: sizes of clusters of cases, traffic data, and imported cases outside the ME region. Methods used account for the limited information available and reporting inaccuracies. Our aim is to complete early findings on MERS-CoV epidemic by focusing on the virus transmissibility from human-to-human, its possible changes in time, the expected number of cases in the region, and the public health threat for other geographical areas based on case importation.

**Methods**

**Analytic overview.** The integrative approach we use is based on a combined maximum likelihood analysis to jointly estimate the reproductive number $R$ and the daily rate $p_{sp}$ of sporadic introduction of the virus in the population through zoonotic/environmental transmissions [10]. The integrative approach builds on two aspects of the currently reported outbreak – the distribution of cluster sizes, providing information on $R$ (Method 1), and the number of imported cases in countries out of the source region providing information on $R$ and $p_{sp}$ based on the fit of a stochastic spatial metapopulation model integrating aviation data worldwide (Method 2).

**Method 1.** We considered laboratory-confirmed cases in the ME region (Jordan, Qatar, Kingdom of Saudi Arabia, United Arab Emirates) reported to WHO as of August 31, 2013 [1] (Table 1). We estimated the reproductive number using the cluster sizes distribution. Several distributions can be used that correspond to different hypotheses regarding the number of secondary cases distribution (offspring distribution) [11]. In particular we considered a Poisson offspring distribution accounting for no overdispersion around a common mean [12] and a geometric offspring distribution assuming a constant rate of transmission during an exponentially distributed infectious period [13].

As cluster sizes may be biased downwards by incomplete observation, we allowed for uncertainty by assuming that each case in a cluster would go unobserved with probability $p_{cl}$ during investigation ($p_{cl} = 0$ representing no missed cases). This corresponds to the following distribution for reported cluster sizes:

$$P(O = k | R, p_{cl}, O \geq 1) = \frac{\sum_{j \geq k} P(S=j|R) \binom{j}{k} p_{cl}^{j-k}(1-p_{cl})^k}{1 - P(O=0|R,p_{cl})}, \quad (1)$$

where $O$ is the observed size of the cluster, $S$ its real size ($O \leq S$), $P(O = 0|R, p_{cl}) = \sum_{j \geq 1} P(S = j|R) p_{cl}^j$, and $P(S|R)$ is the offspring distribution discussed above.

Eventually, the likelihood was computed as $\mathcal{L}_1(R, p_{cl}; \{o_i\}) = (\sum_i o_i)! \prod_i \frac{1}{o_i!} P(O_i = o_i | R, p_{cl}, O \geq 1)^{o_i}$ over a bi-dimensional grid of $\{R, p_{cl}\}$ values.

We performed a sensitivity analysis by considering: (i) the complete Jordan cluster including 8 more cases (size=10) identified through a retrospective serology study carried out on 124 individuals [4,14]; (ii) all laboratory-confirmed cases (108) reported worldwide to WHO [1]; (iii) laboratory-confirmed cases in the ME region up to May 31, 2013 [1]. The corresponding cluster size data are reported in Table 1.

**Method 2.** Due to large concern around the ongoing outbreak and enhanced surveillance following the WHO guidelines for patients returning from the affected area, the detection of probable cases imported in countries out of the ME region is expected to be more complete than in the source region where primary cases may have gone undetected. Another source of information to estimate the reproductive number, discounting possible notification/surveillance biases in the source region, is therefore provided by the importation of cases in newly affected countries [15,16]. We thus modify and extend a method already used for the estimation of the seasonal transmission potential of the 2009 H1N1 pandemic based on the calibration of a global epidemic and mobility model (GLEAM) [17,18] to the chronology data of the pandemic invasion [16], by accounting for the different transmission scenarios in the MERS-CoV source region. The method concurrently allows the estimation of the incidence of infection from sporadic cases in the region, and therefore provides a measure of possible underascertainment of cases.

GLEAM is based on a spatially structured metapopulation approach comprising 3,362 subpopulations in 220 countries in the world coupled through mobility connections. The model is informed with high-resolution demographic data for 6 billion individuals and multi-scale mobility data including the full air traffic database from the International Air Transport Association (IATA) and short-range ground mobility obtained from national commuting data [18]. The infection dynamics takes place within each subpopulation and assumes a modified SEIR compartmentalization (susceptible, exposed, infectious, recovered individuals) [19] to account for different transmission scenarios in the ME region [10]: (i) introduction of sporadic infections from zoonotic/environmental transmission with a daily rate $p_{sp}$ and a uniform spatial distribution in the source region; (ii) modified human-to-human transmissibility with respect to standard homogeneous mixing in all subpopulations of the model to allow for large variations in the number of secondary cases produced by a given primary case. Epidemiological parameters for the compartmental model were based on the estimates obtained from the analysis of the outbreak data including 22 cases at a health-care facility in Al-Ahsa in the Kingdom of Saudi Arabia [6], namely average latency period of 5.2 days and generation time of 7.6 days.

The daily rate $p_{sp}$ of sporadic cases emergence in the ME region and the reproductive number $R$ are the free parameters of the model. For each set of values of these two parameters, GLEAM allows the generation of stochastic numerical realizations of the MERS-CoV outbreak simulating the local epidemic in the source region and the possibility of international dissemination through mobility processes entirely based on real data. We thus generate with a Monte Carlo procedure the probability distribution $P_i(n_i)$ of the number $n_i$ of imported MERS-CoV cases in country $i$ out of the seed region as of August 31, 2013 ($4 \times 10^3$ stochastic realizations for each point ($R, p_{sp}$) of the space of parameters). Being all independent importation events, we can define a likelihood function $\mathcal{L}_2(R, p_{sp}; \{n_j^*\}) = \prod_j P_j(n_j^*)$, where $n_j^*$ is the empirically observed number of imported cases per country (see schematic example in Figure 1). We further restrict our analysis to countries of Western Europe, to focus on a region where respiratory diseases surveillance is homogeneous and with a high sensitivity to detect importations. These data consisted therefore in: $n_j^* = 1$ in the United Kingdom, Germany, France and Italy, and $n_j^* = 0$ for all other countries $j$ in the Western European region (Figure 1). We estimated the log-likelihood over a bi-dimensional grid of $(R, p_{sp})$ values and used bivariate linear interpolation over a refined grid.

Given the unknown geographical extension of the source of the MERS-CoV infection in the ME region and its reservoir, we performed a sensitivity analysis by considering: (i) an extended definition of source region to neighboring countries, thus additionally including Bahrain, Iraq, Iran, Israel, Kuwait, Lebanon, Oman, Palestinian territories, Syria, and Yemen (extended source region), based on travel recommendations [20]; (ii) a restricted source region localized only in the Kingdom of Saudi Arabia, i.e. the country in the ME region that reported the largest number of cases; (iii) a variation in the time of the initial emergence of the virus with transmission to humans, assuming that sporadic cases may be introduced up to two months before the known initial cases (Jordan cluster, April 2012 [4]), to allow for lack of identification or detection prior to the Jordan cluster.

**Integrative approach combining Methods 1 & 2.** Methods 1 and 2 were jointly combined in the following integrated likelihood function: $\mathcal{L}(R, p_{sp}, p_{cl}|\{o_i\}, \{n_j^*\}) = \mathcal{L}_1(R, p_{cl}|\{o_i\})\mathcal{L}_2(R, p_{sp}|\{n_j^*\})$, owing to the independence of the two observed processes (cluster sizes and importations). Maximum likelihood estimates were computed over the 3-dimensional grid $(R, p_{sp}, p_{cl})$ and the deviance $D(R, p_{sp}) = -2(\log \mathcal{L}(R, p_{sp}) - \max(\log \mathcal{L}))$ was used to measure distance from the best fit. Associated confidence intervals were obtained by profiling the deviance in the 3D space [21]. It is important to note that such estimates cannot be derived from the maximum likelihood analysis of each Method considered separately, nor conditionally one to the other, and the full computation of $\mathcal{L}(R, p_{sp}, p_{cl}|\{o_i\}, \{n_j^*\})$ needs to be considered. In this respect, our integrative approach represents a substantial advance with respect to prior work based on the analysis of cluster data only [9].

**Air traffic data analysis.** We additionally analyzed the air traffic data integrated into GLEAM to evaluate the traffic capacity of the airports in the ME region and to assess the importation risk of the countries belonging to other regions than Western Europe.

**Results**

The integrated analysis led to a $R$ value equal to 0.50 (95% CI 0.30–0.77) and daily rate $p_{sp}$ of MERS-CoV introductions into the human population in the ME region equal to 0.28 (0.12–0.85) (Table 2). These best estimates were obtained considering a geometric offspring distribution, yielding higher maximum likelihood values in the analysis. The corresponding best estimate for uncertainty in the cluster distribution suggests a consistent fraction of cases missed in cluster investigations ($p_{cl}$=0.35, 95% CI 0–0.85), but little impact of inaccuracies in reported cluster size on the estimates of the other parameters.

The estimated daily rate $p_{sp}$ of sporadic cases in the ME region can be compared to the observed value ($p_{sp}^* = 0.116$), computed based on 60 sporadic cases reported between April 2012 and August 31, 2013 in the region, including 43 sporadic cases and 13 cluster index cases in the ME region (assuming that each cluster is originated by a single index case), as well as 4 laboratory-confirmed imported cases out of the region [1,3,5]. This yields that the true number of cases from zoonotic/environmental transmission might be between 1.03– and 7.32–fold the observed number.

Our estimates are very robust against the addition of cases out of the ME region to the distribution of clusters sizes ($R$ =0.50, 0.31–0.77; no change for $p_{sp}$). An increase in $R$, though with limited change in the associated confidence interval, is obtained if we include the full Jordan cluster of April 2012 by considering also cases retrospectively confirmed by serology ($R$ =0.65, 0.34–0.80); no variations are obtained in the confidence interval of the estimated daily rate of sporadic cases in the region (Table 2).

Similar results for the reproductive number are obtained when we consider variations in the geographic definition of the MERS-CoV source region. Extending it to its neighboring countries does not affect the estimated basic reproductive number and associated confidence interval ($R$ =0.50, 0.31–0.76), but lowers the value of the daily rate of sporadic cases ($p_{sp}$ =0.14, 0.05–0.38). If we assume that the source region is instead restricted to the Kingdom of Saudi Arabia, a substantial increase in $p_{sp}$ is obtained (4.73, 2.32–15.37), with increase in the reproductive number and unaltered confidence interval ($R$ =0.60, 0.30–0.76).

No variation in $p_{sp}$ was observed in testing a different hypothesis on the offspring distribution, whereas an increase for the best estimate of $R$ was found (0.69 in the Poisson case vs. 0.50 in the baseline, with similar CIs, Table 2).

Larger CIs but no significant variation in the parameters' estimates were observed by considering empirical data up to the end of May (Table 2).

Analyses of traffic data expose large traffic fluxes towards the continents of Asia, Europe and Africa (Figure 3) from the ME region. In addition to 6 neighboring countries of the ME region, 7 were found in South Asia that belong to the set of the first 20 countries with highest traffic from the ME region, 5 in Europe (among which the 4 countries reporting importation of cases from the affected area), and 2 in Africa.

**Discussion**

Results of our integrative modeling approach suggest the occurrence of a subcritical MERS-CoV epidemic in the ME region, as quantified by a reproductive number smaller than 1. The outbreak is not able to generate a self-sustaining epidemic in humans, and sporadic cases from zoonotic/environmental transmission are expected to represent a large fraction of the total size of the epidemic.

The estimated confidence interval for the reproductive number is found to be very stable across changes in the data interpretation. In all cases, considering data up to August 31, we found that it is highly unlikely (<5% probability) to have a MERS-CoV outbreak with $R$ above 0.80 or below 0.30. The variation of the best estimate from the baseline case ($R$ =0.50) to the various scenarios explored as sensitivity analysis (up to $R$=0.69) is explained by the presence of a large region in the parameter space ($R, p_{sp}$) where the likelihood function shows small variation around its maximum value (darker red area in Figure 2). This is likely induced by the limited data available not allowing us to narrow down the confidence intervals of the estimates.

The analysis based on the integration of two independent methods allows us to provide an estimate for the daily rate of introductions of MERS-CoV infections in human population in the ME region, in addition to the estimate for the reproductive number. The estimated 95% CI in the baseline scenario (0.12–0.85) compared to the observed value (0.116) suggests a negligible to significant underascertainment rate for zoonotic/environmental transmissions (1.03–7.32 times the reported sporadic cases), indicating that notified sporadic cases likely represent a substantial proportion of the total, but improved surveillance in the region including serological surveys around cases is needed. Since evidence for mild illness, as well as for a wide spectrum of clinical disease, was observed [3,6], our findings are compatible with an underascertainment rate for zoonotic/environmental transmissions that may be due to a selection bias towards more severe cases, where patients having mild illnesses or asymptomatic infections may go undetected [22].

Our estimates for the reproductive number are consistent with the results of Ref. [9] – the only study to date reporting results on interhuman transmissibility – thus further confirming the robustness of our epidemic assessment. Our work presents however substantial differences in the methodology and in its achievable predictions that we discuss in the following.

One major difference is that our integrative approach also allows the quantification of sporadic cases underascertainment through the estimate of the rate of introduction of sporadic cases in the ME region, combined with the estimate for the reproductive number. In [9], daily introductions are simply calculated on the basis of the two assumed scenarios for the transmission trees, i.e. from the assumed number of index cases among the reported data. Our procedure instead makes no assumption on the completeness of reported data, or on the local transmission trees, and relies on alternative data sources (case importations) to estimate the number of sporadic cases in the region.

The cluster data analysis of Method 1 relies on the assumption that each cluster is the result of human-to-human transmission starting from a single index case. While we allow for uncertainty in case detection in the close contact investigation, we do not consider the possibility of co-exposure of epidemiologically linked cases to the same source of zoonotic/environmental infection, differently from [9]. In absence of any knowledge about the virus path of infection to humans and with insufficient data from epidemiological investigations to reliably reconstruct transmission trees within clusters, we chose a worst-case assumption for the transmissibility of the virus. This may lead to overestimating the reproductive number, however not affecting our conclusion on the subcritical nature of the current MERS-CoV epidemic. In addition, such assumption does not affect our analysis based on case importations (Method 2) that aims to estimate the size of the epidemic in the affected region, in that it assumes no knowledge on the local transmissions, independently of their type (whether human-to-human or zoonotic/environmental).

We considered the cluster analysis (Method 1) restricted to the ME region as we assumed a rather homogeneous implementation of control measures around cases that may be different from the one put in place in affected countries experiencing importation of cases, mainly due to the additional available knowledge of travel history associated to imported cases. The extension of the analysis to all MERS-CoV clusters of laboratory-confirmed cases reported to WHO did not alter our estimates.

The spatial component of our approach represents another difference with previous work. If we assume that the source of MERS-CoV infection is restricted to the Kingdom of Saudi Arabia, where the majority of cases has been observed, our estimates indicate that a much larger number of sporadic cases in the area would be needed to sustain the observed importation of cases in Western Europe (4.73 daily introductions of sporadic cases vs. 0.28 in the baseline). The biggest airports, handling the vast majority of the international air traffic of the region, are indeed mainly localized in the United Arab Emirates, Qatar, and Jordan (panel A of Figure 3). This strongly reduces the traffic capacity of the ME region, as well as the corresponding likelihood of exporting cases out of the region, once the restricted hypothesis on the seeding area is considered. An analysis based on cluster data only would remain unchanged and would not be able to detect important variations in the epidemic size estimation. This result further indicates the relevance of air travel in the epidemic assessment and therefore the need for an integrative approach also based on mobility and space. It also implies that improving the knowledge of the geographic extent of the seed region is critical, along with the identification of the virus' path of transmission to humans.

Changes in time of countries' public health actions for surveillance and control of the MERS-CoV epidemic certainly occurred in response to the increasingly available emerging evidence and the higher awareness of the disease, however data is too scarce to provide estimates of the reproductive number $R(t)$ as a function of time. Here we assumed a constant $R$ for the period under study, with the underlying assumption of a constant and homogeneous implementation of intervention measures in the region.

Other possible factors leading to variations of the current situation may be pathogen- and hosts-related. Evolution of the MERS-CoV virus to adapt to humans and reach sustainable and efficient human-to-human transmission represents a potential future scenario, as it happened for SARS [23]. We tested for possible variations of $R$ and $p_{sp}$ estimates by comparing two different points in time (end of May and end of August 2013) and found no significant change, except for a reduction of the CIs following a larger dataset available for the estimation. Given the available data, this result seems to indicate that no variation in the rate of introductions and in the transmissibility of the virus has occurred in the last 3 months that may point to differences in the transmission to humans or to viral adaptations to human hosts. On the other hand, if compared to the reported number of sporadic cases at the two dates, our $p_{sp}$ estimates are in favor of an increase in sporadic case ascertainment as the estimated underascertainment rate has decreased in this time period from 2.2–17.5 considering notified laboratory-confirmed cases up to May 31 to 1.03–7.32 up to August 31. Further data to update the integrative approach will contribute to provide a continuous assessment of the outbreak in case an evolving situation is suspected.

Other events that are related to human movements and mixing may as well alter the assessed scenario. Vast international mass gatherings to be taking place in the Kingdom of Saudi Arabia in the upcoming months are expected to bring large number of pilgrims to the affected area, with increased rates of local mixing that may favor the transmission of the virus, followed by a potential amplification of its international dissemination due to the return of pilgrims to their own countries [24]. It should be noted however that the Hajj 2012 occurred with an ongoing MERS-CoV outbreak in the ME region, and that MERS-CoV was absent among French pilgrims screened prior to returning to France after their participation to the pilgrimage [25]. Similarly, no increase in case notification occurred following the Umra pilgrimage in July 2013. Additional studies in travelers next to enhanced local surveillance in the region and guidance to local authorities would help to assess and control possible changes in the virus transmission with respect to last year's experience [26].

Air travel clearly represents the main mean for global spatial spread of infectious disease epidemics in the modern world, as it was previously experienced with SARS and the 2009 H1N1 pandemic [15,16,27-29]. Besides seasonal variations due to specific events (e.g. mass gatherings) or in/out flows of expats for seasonal jobs, a potential emerging pandemic in the ME area would constitute a very high risk for considerably rapid and wide international spread. The ME area indeed covers a central role in connecting different regions of the world and has witnessed a dramatic increase of traffic growth in the last decade (153% of relative increase in the Kingdom of Saudi Arabia in 2002-2011, 240% in Jordan, 408% in Qatar, 512% in the United Arab Emirates, against a global relative increase of 168%) [30]. Analysis of the air traffic data integrated into GLEAM shows that large fluctuations are observed in the repartition of travel flows on direct connections out of the ME region, suggesting that other countries than the ones already affected are at high risk of MERS-CoV importation through infected passengers, in particular in the Southern regions of Asia (Figure 3). Similar results were also reported in Ref. [24]. Should the outbreak evolve in a self-sustained epidemic, such risk assessment analyses cannot rely on travel data only and would require the full integration of the air travel data with an epidemic model, as in GLEAM, to explicitly simulate the evolving epidemic, estimate importation likelihood [28] and provide predictions for future stages of the epidemic [16,31].

With a subcritical epidemic in the ME region associated with a large potential for international dissemination, priority for the epidemic control should be given to the identification of the transmission of infection to humans to limit sporadic cases, to the reduction of human-to-human transmission through rapid case identification and isolation, and to the enhancement of surveillance systems in those countries that are at a higher risk of importation because of travel flows to/from the affected area.

**Acknowledgments**

The authors would like to thank J-C Desenclos and S Deuffic-Burban for useful interactions and comments.

**Funding statement**

This work is partly supported by the ERC Ideas contract no. ERC-2007-Stg204863 (EpiFor) to ChP and VC; the EC-Health contract no. 278433 (PREDEMICS) to ChP and VC; the ANR contract no. ANR-12-MONU-0018 (HARMSFLU) to DLB and VC. The funders had no role in study design, data collection and analysis, decision to publish, or preparation of the manuscript.

# Tables

**Table 1.** Cluster size data.

| size of cluster | baseline (all lab-confirmed cases in the ME region) | considering the full Jordan April 2012 cluster (including probable cases) | considering all lab-confirmed cases worldwide | considering data up to May 31, 2013 |
|---|---|---|---|---|
| sporadic case | 42 | 42 | 44 | 16 |
| 2 | 8 | 7 | 10 | 2 |
| 3 | 2 | 2 | 4 | 1 |
| 5 | 2 | 2 | 2 | - |
| 10 | - | 1 | - | - |
| 22 | 1 | 1 | 1 | 1 |

**Table 2.** Best estimate values for the reproductive number $R$ and the daily rate $p_{sp}$ of sporadic cases emergence due to zoonotic/environmental transmissions (and corresponding 95% CI). Results for the baseline and for the scenarios of the sensitivity analysis are obtained assuming a geometric offspring distribution for the analysis of cluster data and for the best estimate of the uncertainty parameter, $p_{cl}$. The last line refers to the baseline scenario considering a Poisson offspring distribution for the analysis of cluster data.

| | $R$ | $p_{sp}$ |
|---|---|---|
| baseline | 0.5 (0.30–0.77) | 0.28 (0.12–0.85) |
| considering the full Jordan April 2012 cluster (including probable cases) | 0.65 (0.34–0.80) | 0.28 (0.12–0.83) |
| considering the extended source region | 0.50 (0.31–0.76) | 0.14 (0.05–0.38) [a] |
| considering the restricted source region (limited to the Kingdom of Saudi Arabia) | 0.60 (0.30–0.76) | 4.73 (2.32–15.37) [a] |
| considering data up to May 31, 2013 | 0.54 (0.34–0.90) | 0.43 (0.12–0.95) |
| considering a Poisson offspring distribution | 0.69 (0.34–0.79) | 0.28 (0.12–00.71) |

[a] to be comparable with the other estimates, this value has been rescaled to take into account the change of population size of the seed region; it thus represents the daily rate of sporadic cases scaled to the ME region.

# Figures

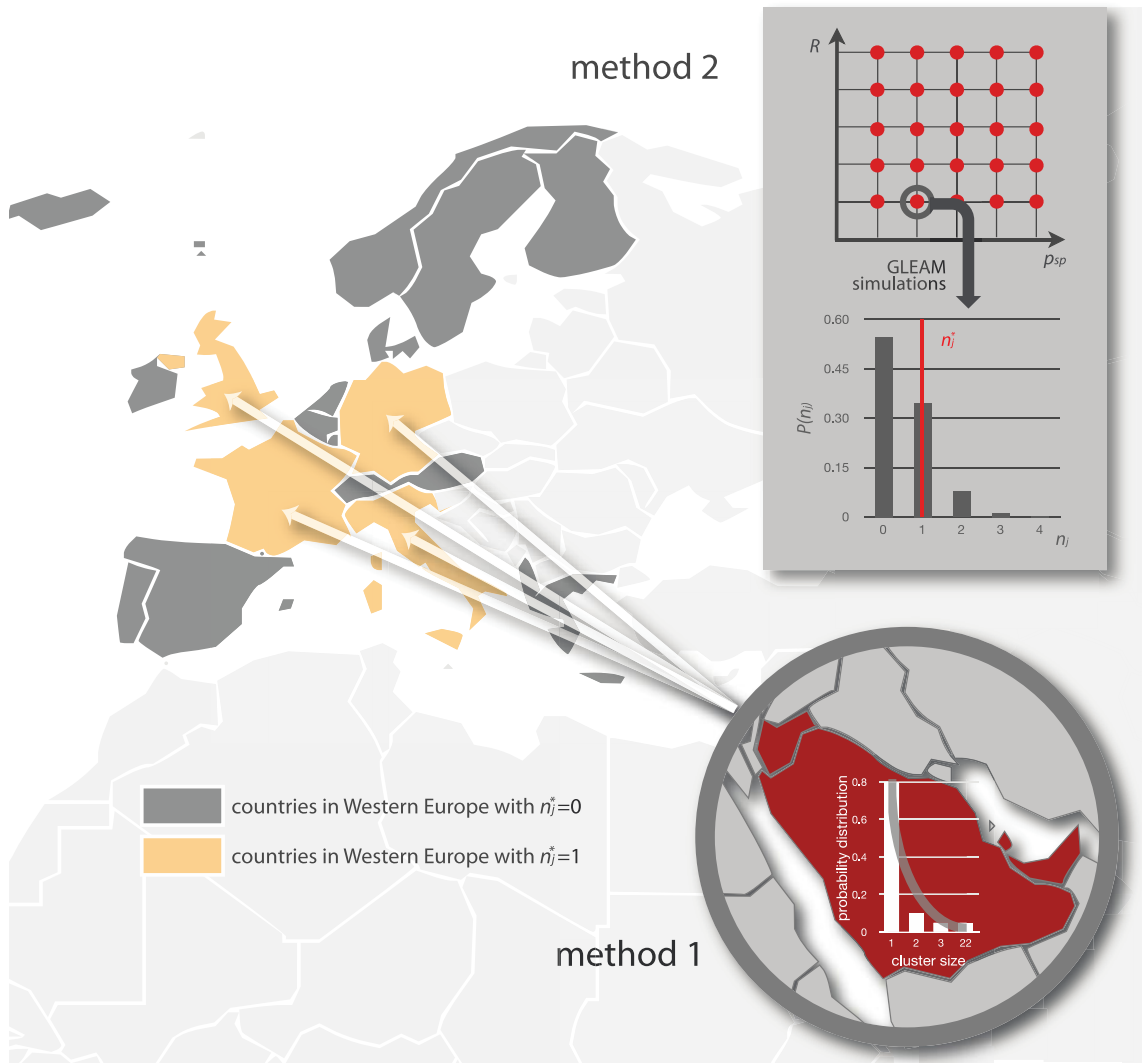

**Figure 1. Schematic representation of the integrative maximum likelihood approach.** Method 1 (bottom circle) is based on the maximum likelihood analysis of cluster size distribution obtained from laboratory-confirmed cases in the ME region (countries in red in the zoomed area). Method 2 (top panel) is based on the maximum likelihood analysis on data on case importations in countries in Western Europe, as schematically indicated on the map. For each point in the parameter space $(R, p_{sp})$ we run 4,000 stochastic GLEAM simulations from the same initial conditions and parameterized as described in the main text. With each run providing the simulated number of imported cases $n_j$ for a given country $j$, we can compare the resulting simulated probability distribution of $n_j$ with the observed value $n_j^*$ for that country and compute a likelihood function for all countries in Western Europe.

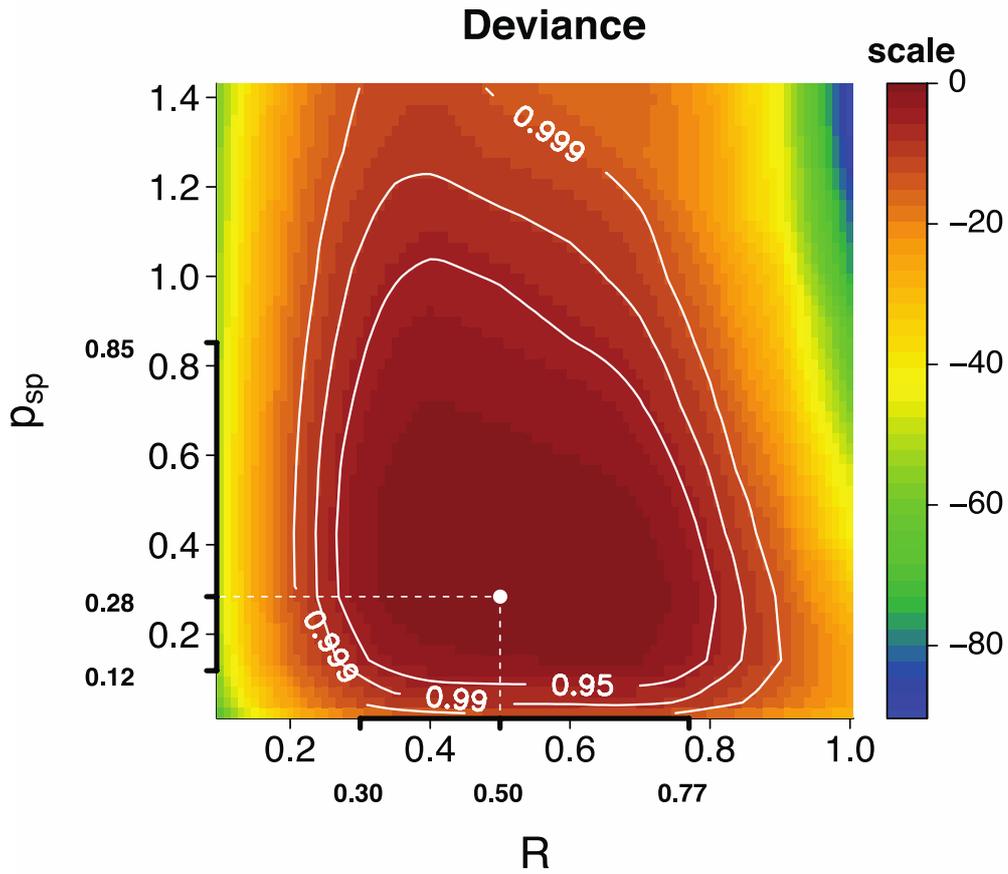

**Figure 2. Heatmap of deviance values vs. reproductive number $R$ and daily rate $p_{sp}$ of sporadic cases.** Deviance was calculated as $D(R, p_{sp}) = -2\big(\log \mathcal{L}(R, p_{sp}) - \max(\log \mathcal{L})\big)$ using the profiled log likelihood (for each pair $(R, p_{sp})$, the uncertainty parameter $p_{cl}$ in cluster size distribution maximizing the log-likelihood $\mathcal{L}$ was chosen). Vertical and horizontal dashed lines show the maximum likelihood values for $R$ and $p_{sp}$, respectively. Solid white curves contour the deviance-based confidence regions of levels 95%, 99%, and 99.9%. The 95% profiled confidence intervals for $R$ and $p_{sp}$ are highlighted in bold on the axes.

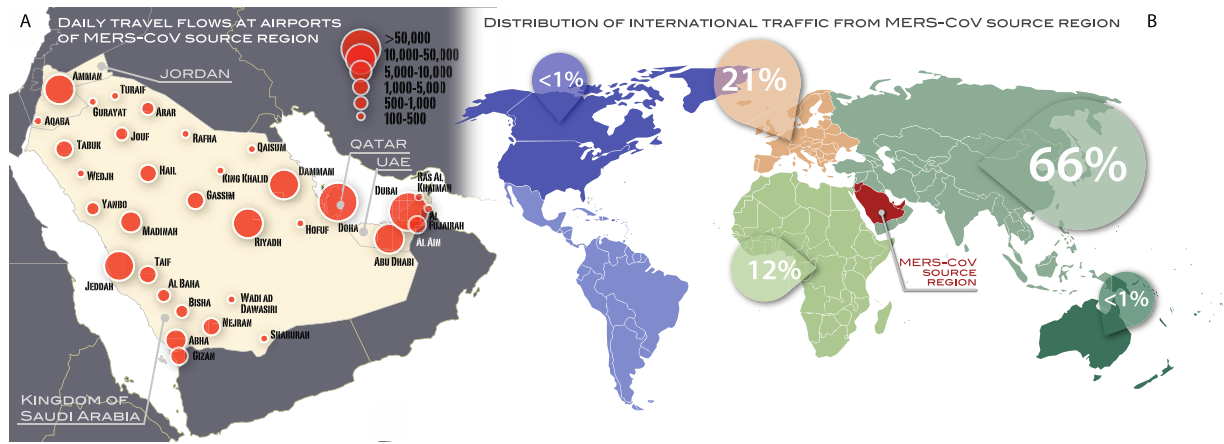

**Figure 3. Air traffic capacity of the MERS-CoV source region and its international destinations.** Airports in the MERS-CoV source region (Saudi Arabia, Jordan, Qatar, United Arab Emirates) are represented with a circle proportional to the daily traffic they handle (panel A). Their international traffic, out of the region, is broken down by continent of destination (panel B). Breakdown by country for the first 20 countries with highest traffic from the ME region: India (11.7%), Bahrain (8.7%), Pakistan (8.6%), United Kingdom (8.4%), Oman (5.8%), Egypt (5.2%), Kuwait (4.3%), Iran (3.6%), Germany (3.5%), Lebanon (2.9%), Bangladesh (2.8%), Thailand (2.5%), Sri Lanka (2.3%), Singapore (2.1%), Syria (2.0%), France (2.0%), Kenya (1.6%), Italy (1.5%), Malaysia (1.4%), Switzerland (1.4%). Statistics are obtained from the IATA air traffic data integrated into the model GLEAM [18].